\documentstyle[aps]{revtex}

\input epsf

\begin{document}


\draft

\preprint{LAEFF-96/14, gr-qc/9606072}

\title{Charge scaling and universality in critical collapse}

\author{Carsten Gundlach and Jos\'e M. Mart\'\i n-Garc\'\i a}

\address{Laboratorio de Astrof\'\i sica Espacial y F\'\i sica Fundamental,
Apartado~50727, 28080~Madrid, Spain}

\date{26 June  1996}

\maketitle


\begin{abstract}

Consider any 1-parameter family of initial data such that data with
parameter value $p>p_*$ form black holes, and data with $p<p_*$ do not. As
$p \to p_*$ from above (``critical collapse''), the black hole mass scales
as $M\sim (p-p_*)^\gamma$, where the critical exponent $\gamma$ is the same
for all such families of initial data.  So far critical collapse has been
investigated only for initial data with zero charge and zero angular
momentum. Here we allow for $U(1)$ charge.  In scalar electrodynamics
coupled to gravity, with action $R + |(\partial + iqA)\phi|^2 + F^2$, we
consider initial data with spherical symmetry and nonvanishing charge. From
dimensional analysis and a previous calculation of Lyapunov exponents, we
predict that in critical collapse the black hole mass scales as $M\sim
(p-p_*)^\gamma$, and the black hole charge as $Q \sim (p-p_*)^\delta$, with
$\gamma = 0.374 \pm 0.001$ (as for the real scalar field) and $\delta =
0.883 \pm 0.007$. We conjecture that, where there is no mass gap, this
behavior generalizes to other charged matter models, with $\delta \ge 2
\gamma$. We suggest the existence of universality classes with respect to
parameters such as $q$.

\end{abstract}



\section{Introduction}

Consider any 1-parameter family of initial data for general relativity
such that data with parameter value $p>p_*$ form black holes, and data
with $p<p_*$ do not. Recently it has been discovered that as $p \to
p_*$ from above (``critical collapse''), and unless there is a mass
gap, the black hole mass scales as $M\sim (p-p_*)^\gamma$, where the
critical exponent $\gamma$ is the same for all such families of
initial data. 

This behavior was first discovered for a real scalar field in spherical
symmetry \cite{Chop}, has been observed in a variety of other matter
models \cite{EvCol,HHS,ChopLieb,EYM}, and is strongly expected to occur in
two larger classes \cite{Maison,HE3}, all in spherical symmetry. Critical
phenomena were also found in axisymmetric pure gravity
\cite{AbrEv}. Historically the second case to be discovered, it has so far
remained alone in going beyond spherical symmetry, and in being vacuum. No
counter-examples have been found, except that when the field equations
allow for a metastable static solution this solution creates a mass gap for
certain initial data regimes \cite{EYM}.

The critical scaling of the mass can now be considered as understood
\cite{EvCol,Koike,G2} in terms of general relativity as a dynamical system,
and of Lyapunov exponents, at least in spherical symmetry.  For reasons of
space we assume here that the reader is already familiar with these
arguments. Our notation here is compatible with \cite{G2}. An introductory
review is \cite{Bizon}.

Black holes are characterized by mass $M$, $U(1)$ charge $Q$ and angular
momentum $L$, but so far critical collapse has been investigated only for
initial data with zero charge and zero angular momentum. (Linear
perturbations of extremal charged black holes ($Q=M$) were studied in
\cite{Traschen}.) Here, as a first step towards the generic situation, we
allow for charge in the initial data and hence in the resulting black
hole. We keep the restriction to spherical symmetry (and therefore exclude
angular momentum).


\section{Scalar electrodynamics}

As a matter model with $U(1)$ charge, we choose scalar electrodynamics
coupled to gravity. This has the advantage of being a simple field theory,
and of allowing arbitrary and independent initial distributions of
energy-momentum and charge. Furthermore it is a generalization of two other
models in which critical collapse has already been studied, namely the real
and complex massless scalar fields (without electromagnetism).  The action,
in units where $c=2\pi G=1$, is
\begin{equation}
S = \int \sqrt{g} \left[ {1\over8} R -{1\over2} g^{ab}
\left(\nabla_a + iqA_a\right)\phi \, \left(\nabla_b -
iqA_b\right)\phi^* -{1\over 4} g^{ab} g^{cd} F_{ac} F_{bd} \right],
\end{equation}
where $F_{ab}\equiv \nabla_a A_b - \nabla_b A_a$. We restrict to spherical symmetry.  For the metric we choose
\begin{equation}
ds^2=-\alpha(r,t)^2 \, dt^2 + a(r,t)^2 \, dr^2 + r^2 \, d\Omega^2
\end{equation}
with remaining gauge condition $\alpha=1$ and regularity condition $a=1$
(or $a_{,r}=0$) at the origin $r=0$ (before a singularity forms there). For
the electromagnetic field we choose the gauge $A_r\equiv0$, and $A_t=0$ at
$r=0$.

In order to use dimensional analysis and scaling arguments later, we need
the field equations in first-order and dimensionless form.  For this
purpose we introduce the dimensionless fields 
\begin{equation}
X\equiv {r\over a} \phi_{,r}, \quad 
Y\equiv {r\over \alpha} \left(\phi_{,t} + i q A_t \phi\right), \quad
X_\pm \equiv X \pm Y, \quad
A \equiv A_t, \quad
F \equiv {r\over a \alpha} A_{t,r}, \quad
g \equiv {a\over \alpha}.
\end{equation}
With the notation $Z\equiv\{X_+, X_-, \phi, a, g, A, F\}$, we can write
the field equations formally as
\begin{equation} 
\label{dimensionless1}
{\bf E}\left(rZ_{,r}, \, rZ_{,t}, \, Z, \, qr \right) 
= 0,
\end{equation}
where the vector ${\bf E}$ of equations is polynomial in its arguments.
(The matter stress-energy tensor and the matter equations of motions are
manifestly polynomial because they have been derived from a polynomial
matter Lagrangian. The components of the Ricci tensor can be brought into
polynomial form after multiplication by certain metric coefficients.)
Furthermore it is linear in the first two arguments, because the field
equations are now first order.

The arguments of ${\bf E}$ form a complete set of independent dimensionless
quantities.  To make dimensionless quantities out of $q$ and the
derivatives, we have combined them with $r$, taking advantage of the
spherical symmetry. (The Ricci scalar $R$ has dimension $L^{-2}$, and
therefore $F_{ab}F^{ab}$ must have the same dimension. This means that
$A_a$ is dimensionless. As the minimal coupling of matter to
electromagnetism is via $\nabla_a \to \nabla_a + i q A_a$, $q$ has
dimension $L^{-1}$.)

Before giving the equations explicitly, we introduce new dimensionless
coordinates
\begin{equation}
\label{tauzeta}
\tau \equiv \ln \left({t\over r_0}\right), \qquad \zeta \equiv \ln\left({r\over
t}\right) - \xi_0(\tau),
\end{equation}
where $r_0$ is an arbitrary fixed scale, and $\xi_0(\tau)$ is a periodic
function with period $\Delta$. This choice of coordinates is adapted to the
self-similar solution we shall be looking for \cite{G2}. In these
coordinates, discrete self-similarity corresponds to all fields being
periodic in $\tau$. (As a degenerate case, continuous self-similarity
corresponds to all fields being independent of $\tau$.) 

Rewriting the arguments of ${\bf E}$ in equation (\ref{dimensionless1}), we
obtain
\begin{equation} 
\label{dimensionless2}
{\bf E}\left(Z_{,\zeta}, \, e^{\zeta + \xi_0} [Z_{,\tau} - (1
+ \xi_0') Z_{,\zeta}], \, 
Z, \, q r_0 e^{\tau + \zeta + \xi_0} \right) = 0.
\end{equation}
By shifting the origin of $\tau$, one may set $r_0=q^{-1}$ to simplify the
notation, but we want to stress here that $q$ is always accompanied by
$e^\tau$ in the field equations, and vice versa. This will be crucial later
on. 

The complete field equations in these fields and coordinates, written out
explicitly, are
\begin{eqnarray}
\label{matter}
X_{\pm,\zeta} && = {\left[{1\over2}(1-a^2) - a^2 |X_\mp|^2 + a^2
F^2\right] X_\pm - X_\mp  
\pm e^{\zeta + \xi_0} \left[ g X_{\pm,\tau}
+ i q r_0 e^\tau \left(g A X_\pm + a F \phi\right) \right]
\over 1 \pm (1+\xi_0') e^{\zeta + \xi_0} g} 
\\
\label{1}
g_{,\zeta} && = \left(1-a^2+2a^2 F^2\right) g \\
\label{2}
a_{,\zeta} && = {1\over 2} a \left[1-a^2 + a^2 \left(|X_+|^2 + |X_-|^2
+ 2 F^2\right)\right]  \\
\label{constraint}
\phi_{,\zeta} && = a X \\
F_{,\zeta} && = - F + i q r_0 e^{\tau + \zeta + \xi_0}
a {1\over 2}\left(\phi Y^* - \phi^*
Y\right) \\
A_{,\zeta} && = g^{-1} a^2 F \\
a_{,\tau} && = e^{-(\zeta + \xi_0)} {1\over 2} a^3 g^{-1} 
\left(|X_+|^2 - |X_-|^2 \right) 
+ (1+\xi_0') {1\over 2} a \left[1-a^2 + a^2 \left(|X_+|^2 + |X_-|^2
+ 2 F^2\right)\right]  \\
F_{,\tau} && = i q r_0 e^\tau 
a g^{-1} {1\over 2}\left(\phi X^* - \phi^* X\right) 
+ (1+\xi_0') \left[ - F + i q r_0 e^{\tau + \zeta + \xi_0}
a {1\over 2}\left(\phi Y^* - \phi^*
Y\right) \right] \\
\label{phi}
\phi_{,\tau} && = a \left[ (1 + \xi_0') X + g^{-1}
e^{-(\zeta + \xi_0)} Y\right] -i q r_0 e^\tau A \phi
\end{eqnarray}
where $\xi_0'\equiv d\xi_0/d\tau$.
The last three equations, which contain only $\tau$-derivatives, can be
considered as constraints (at constant $\zeta$) which are propagated in
$\zeta$ by the first six equations.


\section{Critical solution}

Critical collapse as we know it is dominated by a ``critical solution''
with two crucial properties: it is an attractor of co-dimension one, thus
serving as an intermediate asymptotic, and it is self-similar, thus linking
the large scale of the initial data with the small scale of the final black
hole. The critical solution acts as an attractor within the hypersurface
which separates black hole data from non-black hole data (the ``critical
surface''). It thus funnels all initial configurations on either side of
the surface into two unique channels; one side goes to a black hole, and
the other to dispersion. This explains the universality of the critical
exponent.  The power-law behavior of the black hole mass follows from the
self-similarity of the critical solution.

Finding a critical solution is a two-step process: one has to find a
self-similar solution, and then check explicitly the number of its unstable
perturbations.


\subsection{Asymptotically self-similar solutions}

Self-similarity corresponds to periodicity of the solution in the
coordinate $\tau$, while the field equations contain explicit factors of
$e^\tau$. Therefore no self-similar solution exists. The physical reason
is that the coupling to electromagnetism introduces the scale $q^{-1}$ into
the action and field equations, thus excluding scale-invariant
solutions. 

Does the absence of an exactly self-similar solution rule out critical
phenomena? No, as we really only need a solution which approaches a
self-similar spacetime as the latter approaches its singularity, that is in
the limit of small spacetime scale. (Any spherical self-similar spacetime
has a curvature singularity, with $R\sim e^{-2\tau}$.) This leads us to the
ansatz
\begin{equation}
\label{scaling}
Z(\zeta,\tau) = \sum_{n=0}^\infty e^{n\tau} Z_n(\zeta,\tau),
\end{equation}
where each $Z_n(\zeta,\tau)$ is periodic in $\tau$.  As $\tau \to -\infty$,
$Z$ is dominated by $Z_0$, which is by assumption periodic in $\tau$.
Therefore the form (\ref{scaling}) guarantees that the solution $Z$ is
asymptotically self-similar as $\tau \to -\infty$. Like a self-similar
solution, it has a singularity, at $(t=0,r=0)$, which corresponds to
$\tau=-\infty$ for any $\zeta$.

Now we argue that there are solutions of this form at least up to some
distance from the singularity. We obtain equations for the $Z_n$ by
substituting the ansatz into the field equations and separating powers of
$e^\tau$. $Z_0$ obeys an equation independent of the other $Z_n$, while all
the other $Z_n$ are coupled to $Z_{n-1}$ and lower. This allows us to
determine the $Z_n$ recursively, starting with $Z_0$. Therefore solutions
$Z$ of the full problem are mapped one-to-one to the solutions $Z_0$ of the
zeroth order problem, which we shall see is simpler.

We know from the study of self-similar solutions that $Z_0$ is locally
uniquely given by a nonlinear eigenvalue problem, arising from the boundary
conditions of (i) periodicity in $\tau$ and (ii) regularity at (a) $r=0$
(for $t\ne0$, before the singularity forms) and (b) the past light cone of
the singularity. By analogy, it is plausible that $Z_1$ exists and is
uniquely determined by the source term $Z_0$ and similar boundary
conditions, and so on for all the $Z_n$.

Does the sum converge? The $Z_n$ are periodic in $\tau$. As
$\zeta\to-\infty$, which corresponds to $r\to 0$ (for $t\ne 0$), the $Z_n$
are bounded by the regularity conditions we impose. $\zeta\to\infty$
corresponds to $t \to 0$ (for $r\ne 0$) and is only a coordinate
singularity. The two examples for $Z_0$ discussed below have been
continued, by a change of coordinates, up to the future light cone, and
have been verified explicitly to be bounded. As the higher $Z_n$ obey
equations similar in type to $Z_0$, we assume that all $Z_n(\zeta,\tau)$
are bounded. As for the growth of the $Z_n$ with $n$, it seems very likely
that it is bounded as $Z_n<Ae^{-\tau_1n}$ for some $\tau_1$, so that the
solution converges for $\tau<\tau_1$. The fact that it does not converge
for all $\tau$ is not crucial to its usefulness. Even when the critical
solution is exactly self-similar and therefore known to exist for all
$\tau$, an actual collapse spacetime is not even approximately self-similar
outside a bounded spacetime region.

What is the meaning of the expansion (\ref{scaling}) in powers of $e^\tau$?
As $e^\tau$ always appears together with $q$ in the field equations, it is
also an expansion in powers of $q$, that is in the coupling of the matter
to electromagnetism. For example, while $A_0$ vanishes, $A_1$ is the
electromagnetic field created by $\phi_0$, and $\phi_2$ is the reaction of
$\phi$ to that electromagnetic field, and so on.  In particular, the
equation for $Z_0$ is obtained by writing $Z_0$ for $Z$ in the field
equations while setting all appearances of $q$ equal to zero. In effect,
$Z_0$ obeys the field equations for a complex scalar field coupled only to
gravity, but not to electromagnetism, which, as we anticipated, is a
simpler problem. Exactly two self-similar solutions of that problem, or
candidates for $Z_0$, are known:

1) The critical solution found by Choptuik \cite{Chop,G1,G2} has no
charge, because the scalar field is real (up to the remaining $U(1)$ gauge
freedom $\phi \to e^{ic}\phi$ with $c$ a constant). This means that all the
$Z_n$ for $n\ge 1$ have vanishing source terms and therefore vanish, or in
other words that the Choptuik solution is already a solution of the full
problem. (Strictly speaking, we have to complete the real Choptuik solution
by a vanishing $A$ and $F$ and a real $\phi$ calculated from
(\ref{phi}). $\phi$ is periodic because in the Choptuik solution $X$ and
$Y$ have vanishing mean value. This is essential, but was not automatically
guaranteed.)

2) The continuously self-similar, charged solution found by Hirschmann
and Eardley \cite{HE1} does have charge. This means that it gives rise to a
solution $Z$ of the full problem, including electromagnetism, only when all
the terms $Z_n$ are added up.


\subsection{Perturbation spectrum}

Now we come to the second criterion for a solution to be a critical
solution, namely the presence of exactly one linear perturbation mode that
is growing as $\tau\to-\infty$. Anticipating the next section, we note that
the general linear perturbations $\delta_i Z$ of $Z$ ($i$ labels the
independent modes) have a similar series form to the background critical
solution $Z$ itself. As for the background expansion(where $Z$ is
completely determined by $Z_0$), $\delta_i Z$ is specified by giving its
leading term $\delta_{i0} Z$. This obeys an equation where (a)
once more the terms with $q$ have been set equal to zero, and where (b) the
background dependence is only through $Z_0$. This means $\delta_{i0}Z$ on
its own is a complete perturbation, in the theory without electromagnetism,
of $Z_0$, which is itself a complete background solution of that theory.
 
The spectrum of the perturbations is determined, in a linear eigenvalue
problem, together with the $\delta_{i0}Z$ (and is not influenced by the
higher $\delta_{in}Z$).  In consequence, the spectrum is the same in the
presence of the coupling to electromagnetism, or in its absence.  We can
therefore recycle earlier work, where the perturbation spectra of the
Choptuik and Hirschmann and Eardley solutions were determined in the theory
without electromagnetism: their spectrum in scalar electrodynamics is just
the same.  As in particular the number of unstable modes is the same, we
immediately conclude that the Choptuik solution is an attractor of
co-dimension one \cite{G2}, that is a genuine critical solution, even in
the full theory with electromagnetism, while the complex solution based on
(but not identical with!) the Hirschmann and Eardley solution has three
unstable modes \cite{HE2}. (To state it once more, its perturbation mode
functions are different in the full theory, but their Lyapunov exponents
are the same.)


\subsection{Uniqueness of the critical solution}

Collapse simulations \cite{privcom} suggest that the Choptuik solution is
in fact the {\it only} critical solution, and a global intermediate
attractor, for the free complex scalar field {\it not} coupled to
electromagnetism. This type of evidence can never be complete, because the
entire phase space can never be probed. It only suggests that any other
other critical solutions have rather small basins of attraction.  As we
have seen, self-similar solutions and their perturbations are mapped
one-to-one between the free complex scalar field and the theory with
electromagnetism. This means that critical solutions are mapped one-to-one
between the theories. Their basins of attractions, however, could be very
different in size, so that the counterpart of a hypothetical critical
solution for the free complex scalar, with a basin of attraction so small
that it would have been overlooked so far, could play an important role in
scalar electrodynamics.

We must also consider the possibility of critical solutions not of the form
(\ref{scaling}). Two such possibilities have occurred to us, but can
be definitely ruled out. 

Any critical solution of the form (\ref{scaling}) describes a situation
where charge becomes less and less important in the final black hole as one
fine-tunes along some one-parameter family of initial data with the aim of
making black holes of ever smaller mass. (This does not exclude {\it
initial data} with an appreciable charge-to-mass radius, but most of that
charge must be radiated away along with most of the mass.) The other
possibility for the charge allowed by the limit $Q<M$ for black holes is
that $Q/M$ approaches a constant as $M\to 0$ in the process of
fine-tuning. This is excluded: The black hole charge $Q$ for a given scalar
field evolution must change sign when $q$ changes sign. As $q$ always
appears in the company of $e^\tau$, $Q$ must be suppressed with respect to
$M$ by at least one power of $e^\tau$ as small scales are approached.

A second possibility (suggested by \cite{EYM}) is that of another critical
solution which is not self-similar but static, giving rise to a mass gap in
the one-parameter families of data. (We come back to this possibility
below, when we consider other matter models with charge.) But massless
scalar electrodynamics has no static solutions, even unstable. ``Charged
boson stars'' exist only if one adds a mass term $m^2\phi^2$ to the action,
with $4\pi G m^2 > q^2$ \cite{Jetzer}.

We come to the surprising conclusion that the Choptuik solution is a
critical solution in scalar electrodynamics, and, as far as we can see, the
only one, even though it has no charge itself. (In particular, the
Hirschmann and Eardley solution, although charged, is not a critical
solution.)

Ongoing work of C. G. shows that the critical solution for spherical
$SU(2)$ Einstein-Yang-Mills collapse \cite{EYM} is of the form
(\ref{scaling}), with all the terms $Z_n$ nonvanishing. This indicates that
if they vanish for $n>0$ in scalar electrodynamics, this is
accidental. Details will be published elsewhere.


\section{Linear perturbations}

We now consider the real critical solution of Choptuik and its linear
perturbations. Because the background is real, these split naturally into
two kinds. The first are perturbations $\delta Z$ with $\delta \phi$ purely
real. In the following we call these the ``real'' perturbations. The
general perturbation of this kind is of the form
\begin{equation}
\label{realpert}
\delta Z_{\text{real}}(\zeta,\tau) = \sum_{i=1}^{\infty} C_i \
e^{\lambda_i\tau}\, \delta_i Z(\zeta,\tau),
\end{equation}
where each $\delta_i Z$ is periodic in $\tau$ with period $\Delta$,
and where the $C_i$ are free parameters. They were already
calculated within the real scalar field model, in \cite{G2}. In
particular there is precisely one eigenvalue $\lambda$ with $\Re
\lambda < 0$ in the spectrum $\{\lambda_i\}$, namely $\lambda_1 \simeq
-2.674$.

The linearized equations for the remaining, not purely real,
perturbations are of the form
\begin{equation}
\delta Z_{,\zeta} = A \ \delta Z + B \ \delta Z_{,\tau}, + e^\tau \, C \
\delta Z,
\end{equation}
where the dimensionless coefficients $A$, $B$ and $C$ are periodic in
$\tau$. The ``real'' perturbations obey the same equations, but the term
proportional to $e^\tau$ (or, as always, proportional to $q$) vanishes for
the real perturbations because they don't carry charge. Its presence means
that the $\delta_i Z$ cannot be periodic in $\tau$. Instead we have
\begin{equation}
\label{imag1}
\delta Z_{\text{imaginary}}(\zeta,\tau) = \sum_{i=1}^{\infty} D_i \
e^{\mu_i\tau}\, \delta_i Z(\zeta,\tau),
\end{equation}
where each independent mode $\delta_i Z$ is now expanded as
\begin{equation}
\label{imag2}
\delta_i Z(\zeta,\tau) = \sum_{n=0}^\infty e^{n\tau} \ \delta_{in}
Z(\zeta,\tau),
\end{equation}
where only the $\delta_{in} Z$ are periodic in $\tau$. This expansion is
exactly analogous to that of the asymptotically self-similar background
solution, with the difference that, due to the linearity, the ansatz
(\ref{imag1},\ref{imag2}) is generic, and (\ref{scaling}) is not.  The
$\delta_{in} Z$ obey the coupled equations
\begin{eqnarray}
\delta_{i0} Z_{,\zeta} && = \left(A + \mu_i B \right) \ \delta_{i0} Z
+ B \ \delta_{i0} Z_{,\tau}, \\ 
\delta_{in} Z_{,\zeta} && = \left(A +
\mu_i B +n B\right) \ \delta_{in} Z + B
\ \delta_{in} Z_{,\tau} + C \ \delta_{in-1} Z, \quad n\ge 1.
\end{eqnarray}
The equation for $\delta_{i0}Z$ describes purely imaginary perturbations of
the scalar field, not coupled to the electromagnetic field or the
metric. It is homogeneous in $\delta_{i0}Z$, and is complemented by
periodic boundary conditions (with period $\Delta$) in $\tau$, and by
regularity conditions at $\zeta=-\infty$ and $\zeta=0$. A solution
exists only for discrete values of $\mu$. From this equation one
obtains the spectrum $\{\mu_i\}$ of perturbations. This equation is in
fact that for the perturbations, around the real solution, of the
complex scalar field model without electromagnetism.  They have
already been considered in \cite{G2}, with the result that $\Re
\mu_i > 0$ for all $i$. 

The terms $\delta_{in} Z$ for $n\ge1$ are determined both by boundary
conditions and by the source terms $\delta_{in-1}Z$. As the value of
$\mu_i$ has already been fixed as an eigenvalue in the equation for
$\delta_{i0}Z$, the boundary conditions admit no homogeneous solution for
$n\ge 1$, and the solution for $n\ge 1$ is proportional to the source term
$\delta_{in-1}Z$.  The electromagnetic field comes in to order $e^\tau$,
that is $q$, because the purely imaginary perturbation $\delta_{i0}\phi$ of
the real background scalar field $\phi_*$ gives rise to a charge
distribution and therefore generates an electromagnetic field. This charge
distribution is ultimately responsible for the charge of the final black
hole.  To order $e^{2\tau}$, or $q^2$, the scalar field perturbations
acquire a real part and in consequence couple to metric perturbations, thus
also creating a gravitational field.  Nevertheless, we shall refer to these
perturbations as the ``imaginary'' perturbations, because to leading order
they are purely imaginary perturbations of the scalar field.


\section{Mass scaling}

For the generic solution close to the real self-similar solution $Z_*$,
we now have the form
\begin{equation}
Z(\zeta,\tau) \simeq Z_*(\zeta,\tau) 
+ \sum_{i=1}^\infty C_i(p) \ e^{\lambda_i \tau} 
\ \delta_i Z(\zeta,\tau)
+ \sum_{i=1}^\infty D_i(p) \ e^{\mu_i \tau} \sum_{n=0}^\infty
e^{n\tau} \ \delta_{in} Z(\zeta,\tau),
\end{equation}
where the first term is the critical solution, and the second and third
terms its ``real'' and ``imaginary'' perturbations, as discussed above.
The amplitudes $C_i$ and $D_i$ of the perturbations depend on the
initial data in general and hence on the parameter $p$ of a given
one-parameter family of initial data in particular.  

As $\tau \to - \infty$, we can neglect all perturbations but the one
growing mode, associated with $\lambda_1$. As we are interested in the
asymptotic behavior of the charge however, we also keep the most slowly
decaying of the imaginary perturbations (which alone carry charge),
associated with $\mu_1$. By definition we obtain the precisely critical
solution for $p=p_*$, and so we must have $C_1(p_*) = 0$. Expanding
$C_1(p)$ and $D_1(p)$ to leading order, we obtain
\begin{equation}
\label{app}
Z(\zeta,\tau) \simeq Z_*(\zeta,\tau) 
+ (p-p_*) \ {\partial C_1(p_*) \over \partial p} \ e^{\lambda_1 \tau} 
\ \delta_1 Z(\zeta,\tau)
+ D_1(p_*) \ e^{\mu_1 \tau} \sum_{n=0}^\infty
e^{n\tau} \ \delta_{1n} Z(\zeta,\tau).
\end{equation}
To keep the notation compact, we define, following \cite{G2}, 
\begin{equation}
\tau_*(p)\equiv - {1\over \lambda_1} \ln \left({p - p_* \over p_*}\right),
\quad 
\tau_1 \equiv - {1\over \lambda_1}
\ln \left[\epsilon^{-1} \  {\partial C_1 \over \partial \ln p}(p_*)
\right],
\quad 
\tau_0 \equiv \tau_1 + \tau_*(p),
\end{equation}
where $\epsilon$ is an arbitrary small constant (independent of $p$). If we
now fix $\tau = \tau_0$ in the approximate solution (\ref{app}), we obtain
a $p$-dependent family of Cauchy data, namely
\begin{equation}
Z_p(r) \equiv Z(\zeta,\tau_0) = Z_*\left(\ln {r\over r_p}, \
\tau_0\right) + \epsilon \ \delta_1 Z\left (\ln {r\over r_p}, \
\tau_0\right) + K(p) \sum_{n=0}^\infty e^{n\tau_0} \ \delta_{1n} Z\left(\ln {r\over
r_p}, \ \tau_0\right)
\end{equation}
where 
\begin{equation}
\label{rp}
r_p \equiv r_0 \ e^{\tau_0+ \xi_0(\tau_0)},
\qquad
K(p) \equiv  D_1(p_*) e^{\mu_1 \tau_0}.
\end{equation}

$K(p)$ is small, even compared to $\epsilon$, if $e^{\tau_0}$ is
sufficiently small, that is, if $(p-p_*)$ is sufficiently small. We
therefore treat the terms proportional to $K(p)$ as a linear perturbation
throughout.

But now we consider the exact, nonlinear evolution of the data $Z_* +
\epsilon \delta_1 Z$, without treating $\epsilon \delta_1 Z$ as a
perturbation any longer. This is necessary because its presence makes a
qualitative difference at late times. If it has one sign, a black hole is
formed. If it has the other, the matter disperses. Because the solution at
late times is no longer even approximately self-similar, we go back to the
coordinates $r$ and $t$.

The data $Z_* + \epsilon \delta_1 Z$ are purely real, and evolve to a
purely real solution, with vanishing electromagnetic field. Consequently,
the equations determining the solution do not contain the term $qr$, and
are scale-invariant. Therefore the entire solution depends on $r_p$ in the
simple way
\begin{equation}
\label{scalable}
Z(r,t) = f(\bar r, \bar t, \tau_0), \quad \hbox{where} \quad 
\bar r \equiv {r \over r_p} \quad \hbox{and} \quad
\bar t \equiv {t-t_p \over r_p},
\end{equation}
where the (irrelevant) shift in $t$ is $t_p\equiv r_0 e^{\tau_0}$.
$f$ obeys the equation
\begin{equation}
{\bf E}\left(\bar r \bar f_{,\bar r}, \, \bar r \bar f_{,\bar t}, \,
f, \, 0\right) = 0,
\end{equation}
with initial data
\begin{equation}
f(\bar r, \bar t = 0, \tau_0) = Z_*\left(\ln \bar r, \tau_0\right) +
\epsilon \ \delta_1 Z\left (\ln \bar r, \tau_0\right).
\end{equation}
In particular we know that the mass of the black hole that forms in
this solution must be a multiple of the underlying scale $r_p$, namely
\begin{equation}
M = r_p \ e^{\mu(\tau_0)},
\end{equation}
where $\mu(\tau)$ is a function that we do not know, but which is
periodic with period $\Delta$ in $\tau$. The black hole mass as a
function of the family of initial data and the parameter value $p$ is
then
\begin{equation}
\label{mass}
M(p) = r_0 \left({p-p_*\over p_*}\right)^{-{1\over \lambda_1}} e^{\tau_1 +
\nu[\tau_*(p) + \tau_1]},
\end{equation}
where $\nu(\tau)\equiv\mu(\tau) + \xi_0(\tau)$ is a universal wiggle
superimposed on the basic power-law behavior, and the constant
$\tau_1$ (defined above) depends on the initial data family.


\section{Charge scaling}

So far we have only repeated \cite{G2} (which itself is a generalization of
\cite{Koike}). Now we consider the effect of the perturbation proportional
to $K(p)$ in the initial data for $f$. This linear perturbation, say
$\delta f$, obeys the linearization of equation (\ref{dimensionless1}),
which is of the form
\begin{equation}
r \ \delta f_{, t} = A  r \ \delta f_{, r} + B \ \delta f +
qr \, C \ \delta f.
\end{equation}
Rewriting this in terms of $\bar r$ and $\bar t$, we obtain (using the
definition (\ref{rp}) of $r_p$)
\begin{equation}
\bar r \ \delta f_{,\bar t} = A \bar r \ \delta f_{,\bar r} + B \ \delta f +
qr_0 e^{\tau_0 + \xi_0(\tau_0)} \ C \bar r \ \delta f,
\end{equation}
where the dimensionless coefficients $A$, $B$ and $C$ are functions only of
$\bar r$ and $\bar t$ (as well as of the parameter $\tau_0$ characterizing
the background solution $f$). To obtain a solution for all small values of
the parameter $e^{\tau_0}$ at once, we expand in powers of it. For the
critical solution plus linear perturbation we then have
\begin{equation}
Z_p(r,t) = f\left(\bar r, \bar t, \tau_0\right) + K(p) \sum_{n=0}^\infty
e^{n\tau_0} \ \delta_n f\left(\bar r, \bar t, \tau_0\right).
\end{equation}
The expansion coefficients $\delta_n f$ obey the equations
\begin{eqnarray}
\bar r\ \delta_0 f_{,\bar t} = && A \bar r \ \delta_0 f_{,\bar r} 
 + B \ \delta_0 f, \\ 
\bar r\ \delta_n f_{,\bar t} = && A \bar r \ \delta_n f_{,\bar r} 
 + B \ \delta_n f + q r_0 e^{\xi_0(\tau_0)}\ C \bar r \ \delta_{n-1} f, \qquad n\ge 1,
\end{eqnarray}
with initial data given by
\begin{equation}
\delta_n f(\bar r, \bar t = 0,\tau_0) = 
 \ \delta_{1n} Z\left(\ln \bar r, 
\tau_0\right).
\end{equation}

The perturbation $\delta f$ gives rise to a perturbation of the black hole
mass $M$, but we ignore this here as a subdominant effect. We are however
interested in the black hole charge $Q$, which {\it only} comes in through
$\delta f$. We do not need to calculate $\delta f$ to see how $Q$
scales. It is sufficient to note that the charge-to-mass ratio is
dimensionless and must be odd in the coupling constant $q$, and hence
$e^{\tau_0}$, and therefore must go as
\begin{equation}
{Q \over M} = K(p) \left[ e^{\tau_0} \left({Q \over M}\right)_1 \! (\tau_0) 
+ e^{3\tau_0} \left({Q \over M}\right)_3 \! (\tau_0) 
+ O\left(e^{5\tau_0}\right) \right].
\end{equation}
Taking only the dominant term, and putting it all together, we obtain
\begin{equation}
Q(p) = M(p) \ K(p) \ e^{\tau_0} \ \left({Q \over M}\right)_1 \! (\tau_0).
\end{equation}
After regrouping terms, we obtain the final result
\begin{equation}
Q(p) = r_0 \ \left({p-p_*\over p_*}\right)^{-{\mu_1+2\over \lambda_1}}
\ e^{\tau_2+\pi[\tau_1+\tau_*(p)]},
\end{equation}
where $\tau_2 \equiv (\mu_1 + 2) \tau_1 + \ln D_1(p_*)$ is a new
family-dependent constant, $\tau_1$ is the same family-dependent
constant as before, and $\pi(\tau) \equiv \nu(\tau) +
\ln \, (Q/M)_1(\tau)$ is a new universal wiggle.

Numerical values for the Lyapunov exponents are $\lambda_1 = - 2.674 \pm
0.009$ \cite{G2} and $\mu_1 = 0.362 \pm 0.012$. $\mu_1$ has been
calculated by the same method \cite{G2} as $\lambda_1$, but converges
somewhat more slowly with decreasing step size, so that the estimated
numerical error in $\mu_1$, and in consequence in $\delta$, is somewhat
larger. We obtain critical exponents $\gamma = - 1 / \lambda_1 = 0.374 \pm
0.001$ for the mass and $\delta = - (\mu_1 + 2) / \lambda_1= 0.883 \pm
0.007$ for the charge.


\section{Other charged matter models}

Given that our arguments did not rely on the exact form of the field
equations, we can generalize them to any matter model in spherical symmetry
where the matter is coupled to electromagnetism only via the
$U(1)$-covariant derivative $D_a \equiv \nabla_a + i q A_a$.  Then it
follows from dimensional analysis that whenever one casts the field
equations in first-order, dimensionless form, they must be of the form
(\ref{dimensionless1}). $Z$ would stand for another set of fields, and
${\bf E}$ need not be a polynomial.  The argument showing that the
electromagnetic interaction can be neglected asymptotically in the strong
field/ small scale regime of the critical solution would go through as
before, even if the critical solution itself carries charge. (Strictly
speaking, we require that a polynomial form of the field equations exists
for our arguments to apply directly, but we hope that this technical
requirement can be relaxed.) Now we need to consider three cases
separately:

(1) The critical solution does not carry charge, only its perturbations
do, which brings in Lyapunov exponents. Let $\lambda_1$ be the one
eigenvalue with $\Re \lambda < 0$. 

(1a) The unstable mode itself carries no charge. This case is similar to
scalar electrodynamics.  Let $\mu_1$ be the eigenvalue associated with
charge which has the smallest real part. We must have $\Re \mu_1 > 0$,
because the critical solution, by definition, has only one unstable
mode. From $\gamma = - 1 / \lambda_1$ and $\delta = - (\mu_1 + 2) /
\lambda_1$ we then find the relation $\delta > 2 \gamma$.

(1b) The unstable mode itself carries charge. In this case, we obtain the
charge, as well as the mass, from the nonlinear evolution of the data $Z_*
+ \epsilon \delta_1 Z$, neglecting {\it all} the other perturbations. The
solution arising from these data is no longer of the simple form
(\ref{scalable}), due to the presence of $qr$ in the equations
(\ref{dimensionless1}). But we can obtain the solution for all small $r_p$
by expanding in $qr_p$, using the fact that $qr=\bar r q r_p$.  The
solution (\ref{scalable}) is now replaced by the more general
\begin{equation}
Z(r,t) = \sum_{n=0}^\infty (qr_p)^n f_n(\bar r, \bar t, \tau_0).
\end{equation}
The mass $M$ is again proportional to $r_p$, but this time only to leading
order in $qr_p$ (or $e^{\tau_0}$). The charge-to-mass ratio is now $Q/M =
(qr_p) (Q/M)_1(\tau_0) + O(qr_p)^3$, without the prefactor $K(p)$, so that
the charge to leading order is simply proportional to $r_p^2$, or $\delta =
2 \gamma$, with $\gamma = - 1/\lambda_1$.

(2) Finally, the critical solution itself may carry charge. Then the higher
terms in the expansion (\ref{scaling}) do not vanish identically. This
would have been the case for example for the solution \cite{HE1}. There the
charge would have appeared to order $e^\tau$ of the background expansion
(\ref{scaling}).  We should still need to consider the data $Z_0 + \epsilon
\delta_1 Z$, where $\delta_1 Z$ is the one growing perturbation (charged or
not), in order to determine $r_p$, the spacetime scale on which the
solution leaves the intermediate attractor. But the electromagnetic field
giving rise to the black hole charge $Q$ is now dominated by $A_1$ (the
component $A$ of $Z_1$), which is once more down a factor of $r_p$ from
$Z_0$, so that $Q$ is again proportional to $r_p^2$, or $\delta = 2
\gamma$.

We must make one proviso, namely that a mass gap may exist in critical
collapse.  Recent work on critical collapse of the Einstein-Yang-Mills
system has illustrated this \cite{EYM}: If the model admits a static or
oscillating solution with precisely one unstable mode (such as the
Bartnik-McKinnon solution in Einstein-Yang-Mills), this solution acts as an
alternative intermediate attractor.  As this intermediate asymptotic is not
self-similar, but instead has a finite mass, it gives rise to a mass gap in
some region of initial data space. Our results then only hold for that part
of initial data space where formation of the static solution is
avoided. For a mass gap to occur when a critical solution also exists, it
is essential that a static or oscillating solution not only exists, but has
exactly one unstable mode. If it was unconditionally stable, there would be
a third possible outcome besides black hole formation and dispersion --
formation of a star. (This is of course the situation in astrophysics.) If
it had more than one unstable mode, it would be a lesser attractor than the
critical solution and would be ``missed'' by almost all one-parameter
families of initial data.


\section{Universality classes and the renormalisation group}

Finally, we note a consequence of our work that does not concern charge.
By exactly the same argument which shows that in the critical solution with
charge electromagnetism can be neglected asymptotically, one can show that
the mass term can be neglected asymptotically for the massive scalar field.
(One only has to replace $qr$ by $mr$ in equation (\ref{dimensionless1})
and the following equations derived from it.) As the critical exponent for
the black hole mass is unchanged by the coupling to electromagnetism, so it
is, by the same argument, when a mass term or a more general polynomial
scalar field self-interaction is added to the action. (This was already
known \cite{Chop2}.)

Our argument generalizes this to {\it any} parameter $m$ of dimension
$(\hbox{length})^{-1}$ in geometrical units $G=c=1$ appearing in a
polynomial form of the field equations. (Note that as we insist on the
equations being polynomial in the fields $Z$ and parameter $m$, we
implicitly also allow parameters of dimension $(\hbox{length})^{-n}$, but
only for positive, integer $n$.)

In the language of critical phenomena, theories with different values of
such a dimensionful parameter form ``universality classes''
\cite{Bizon}. The analogy with critical phenomena in statistical mechanics
seems good enough to use this term deliberately.

The renormalisation group in statistical mechanics acts on the phase space
by blocking degrees of freedom, and a change of scale. If one demands that
the partition function remain invariant, this induces an equivalent action
on (the parameters of) the Hamiltonian \cite{Goldenfeld}.  The equivalents
of the spins and their Hamiltonian in statistical mechanics are the fields
$Z(r)$ and their equations of motion in critical collapse.  The
renormalisation group in critical collapse acts on the phase space by a
time evolution (in $t$) followed by change of scale (in $r$), as $Z(r,t)
\to Z(e^{-\Delta}r, e^{-\Delta}t)$. (One can combine the time evolution and
rescaling into a time evolution in $\tau$, at constant $\zeta$, using the
freedom of lapse and shift in general relativity. The critical solution is
by definition invariant under this transformation.) If one demands that the
rescaled data evolve in the same way, this action induces an equivalent
action on (the parameters of) the field equations.

In the case of scalar electrodynamics, this action is rather simple: In the
field equations, $q$ and $r$ appear together as $qr$. If one changes the
scale as $r \to e^{-\Delta} r$, this has the same effect as $q \to
e^{-\Delta} q$.

Critical phenomena in gravitational collapse correspond to very small
scales (compared to the scale of the initial data and the scale $q^{-1}$),
in contrast to critical phenomena in statistical mechanics, where they
correspond to very large scales (compared to the scale of the microscopic
physics). In the limit of small scales, the parameter $q$ becomes
``irrelevant'', to borrow another term from statistical mechanics.

If there are two (or more) parameters such as $m$ or $q$, universality does
not hold, as $m_1/m_2$ is a dimensionless parameter, which generically has
qualitative effects. In massive scalar electrodynamics, for example, the
value of $|q|/m$ determines if static solutions (``charged Boson stars'')
exist. Universality may be recovered if the two scales are very different,
that is for $m_1/m_2\to 0$ or $m_2/m_1 \to 0$, but this limit need not be
regular.


\section{Conclusions}

We predict that in critical collapse of massless scalar electrodynamics in
spherical symmetry, the black hole mass scales as $M\sim (p-p_*)^\gamma$,
and the black hole charge as $Q \sim (p-p_*)^\delta$, each overlaid with a
universal wiggle, with $\gamma = 0.374 \pm 0.001$ (as for the real scalar
field) and $\delta = 0.883 \pm 0.007$.  

We have gone beyond the restriction to uncharged initial data (and hence
black holes), but have kept the restriction to vanishing angular
momentum. We predict the appearance of a new critical exponent $\delta$,
and give its numerical value. Verification of our predictions in numerical
collapse simulations should be straightforward.

Furthermore, we predict that in spherical critical collapse of matter
models with minimal coupling $\nabla \to \nabla + iqA$ to electromagnetism,
both mass and charge scale as universal power laws (times a universal
wiggle), with $\delta \ge 2 \gamma$, so that the black hole charge always
disappears faster than the black hole mass. Depending on the matter (not in
massless scalar electrodynamics) this behavior may hold only for parts of
the initial data space, with a mass gap in other parts.

We also find that adding terms with a parameter $m$ of dimension
$(\hbox{length})^{-1}$ in units $c=G=1$ (such as a scalar field mass, or
the electromagnetic coupling in the present paper) -- in a form of the
field equations which is polynomial in the fields $Z$ and parameter $m$
-- does not change the critical exponent for the black hole mass. In this
sense models with different values of such a parameter form universality
classes.

{\it Notes added}: After this paper had been submitted, our value of the
critical exponent for the charge was confirmed in numerical collapse
situations by Hod and Piran \cite{HodPiran}. Universality classes in
critical collapse were independently described by Hara, Koike and Adachi
\cite{Haraetal}.


\acknowledgments

We thank Juan P\'erez-Mercader for a critical reading of the manuscript and
suggestions.  C. G. would like to thank Piotr Bizo\'n for discussions on
Einstein-Yang-Mills, and Geoff Simms and Nigel Goldenfeld for suggestions,
especially for expanding section VIII. He was supported by a scholarship of
the Ministry of Education and Science (Spain). J. M. M. was supported by a
scholarship under the 1994 Plan de Formaci\'on de Personal Investigador of
the Comunidad Aut\'onoma de Madrid.



\end{document}